%% file: main.tex
\documentclass[11pt]{article}
\usepackage{subcaption}

\usepackage{fullpage}
\usepackage[utf8]{inputenc} 
\usepackage[T1]{fontenc}    
\usepackage{hyperref}       
\usepackage{url}            
\usepackage{booktabs}       
\usepackage{amsfonts}       
\usepackage{nicefrac}       
\usepackage{microtype}      
\usepackage{graphicx}
\usepackage{caption}
\usepackage{subcaption}
\usepackage{xcolor}
\usepackage{soul}
\usepackage{doi}
\usepackage{authblk}
\usepackage{wrapfig}
\usepackage{amsmath}
\usepackage{epstopdf}
\usepackage{algorithmic}
\usepackage{mathtools}
\usepackage{setspace}
\usepackage[backend=biber,style=numeric, sorting=none]{biblatex}

\DeclareBibliographyDriver{standard}{%
  \printfield{number}\addcomma\space
  \printfield{year}\addcomma\space
  \printfield{title}\addcomma\space
  \printnames{author}.
\finentry
}

\providecommand{\keywords}[1]
{
  \small	
  \textbf{Keywords:} #1
}
\title{Bayesian identification of fibrous insulation thermal conductivity towards design of spacecraft thermal protection systems}

\date{January 20, 2026}

\begin{document}

\author[1]{Alex Alberts}
\author[1]{Akshay Jacob Thomas}
\author[2]{Kamran Daryabeigi}
\author[1]{Ilias Bilionis\footnote{Corresponding author   \href{mailto:ibilion@purdue.edu}{\color{blue}{\texttt{ibilion@purdue.edu}}}}}
\affil[1]{School of Mechanical Engineering, Purdue University, West Lafayette, IN}
\affil[2]{KDT Consulting, Virginia Beach, VA}

\maketitle

\begin{abstract}
    The design of spacecraft thermal protection systems (TPS) requires accurate knowledge of thermal transport properties across wide ranges of temperature and pressure. For fibrous insulation, conventional measurement techniques in laboratory settings are typically limited to temperatures much lower than what is reached in atmosphere entry scenarios. Moreover, it is often the case that only temperature measurements are available, meaning that the thermal conductivity of the insulation must be indirectly inferred as an inverse problem. We propose a Bayesian framework using information field theory (IFT) to reconstruct the thermal conductivity of high-temperature fibrous insulation from sparse experimental data. Under IFT, the conductivity is represented as a Gaussian process, and the physics is enforced via a physics-informed prior over the temperature derived from the heat equation. Bayes's rule produces an infinite-dimensional posterior distribution that quantifies uncertainty about the conductivity which can be evaluated in extrapolation regimes. We apply the method to Opacified Fibrous Insulation with both synthetic and experimental data to reconstruct the thermal conductivity beyond the experimental regime. The inferred conductivities are validated against reference data and then propagated into high-fidelity digital twins of flexible TPS performance under Mars and Earth entry trajectories. The results show that IFT yields accurate predictions with quantified uncertainty, enabling robust TPS sizing in regimes inaccessible to direct measurement.
\end{abstract}


\onehalfspacing

\input{intro}

\input{datageneration}

\input{ift}

\input{TPS}

\input{conclusions}

\paragraph{CRediT authorship contribution statement}
\textbf{Alex Alberts:} methodology, software, validation, writing - original draft, visualization. \textbf{Akshay Jacob Thomas:} methodology, software, data curation, writing - original draft. \textbf{Kamran Daryabeigi:} conceptualization, investigation, resources, writing - original draft, supervision. \textbf{Ilias Bilionis:} conceptualization, resources, writing - review and editing, supervision, funding acquisition.

\paragraph{Funding}
This work was supported by the AFOSR program on materials for extreme environments under grant number FA09950-22-1-0061. 

\printbibliography

\appendix

\section{Eignevalues and eigenfunction of the squared exponential kernel}
\label{apdx:A}

The squared exponential kernel is a special case where the solutions to eq.~(\ref{eqn:Fred}) are known analytically if $p(u) = \mathcal{N}(u|0,\sigma^2)$ is a Gaussian measure~\cite{zhu1997gaussian}.
For $n\in\mathbb{N}$, the eigenvalues $\lambda_n$ and eigenfunctions $\phi_n$ associated with the kernel $s(u,u') = \exp\left(-(u-u')^2/2 \ell^2\right)$ are
$$
\lambda_n = \sqrt{\frac{2a}{A}}B^n
$$
and
$$
\phi_n(x) = \exp\left(-(c-a)u^2\right)H_n\left(\sqrt{2c}u\right).
$$
Here, $H_n(u) = (-1)^n\exp(u^2)\frac{d^n}{du^n}\exp(-u^2)$ is the $n^{\text{th}}$-order Hermite polynomial, $a^{-1} = 4\sigma^2$, $b^{-1} = 2\ell^2$ and
$$
c = \sqrt{a^2 + 2ab}, \quad A= a+b+c, \quad B=b/A.
$$

\section{Temperature and thermal conductivity posteriors across all experiments}
\label{apdx:B}
\begin{figure}[h]
    \centering
    \begin{subfigure}[b]{0.475\textwidth}
        \centering
        \includegraphics[width=\textwidth]{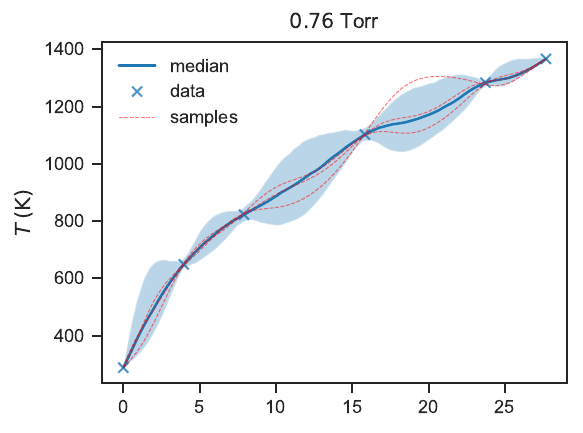}
        \caption{Posterior over $u$ at $0.76$ Torr.}        
    \end{subfigure}
    \begin{subfigure}[b]{0.475\textwidth}
        \centering
        \includegraphics[width=\textwidth]{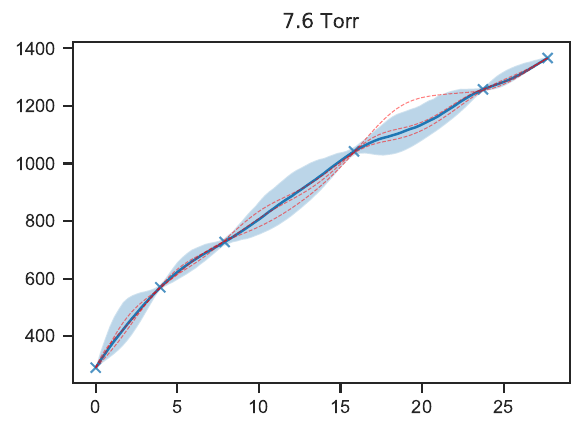}
        \caption{Posterior over $u$ at $7.6$ Torr.}        
    \end{subfigure}
    \hfill
    \begin{subfigure}[b]{0.475\textwidth}
        \centering
        \includegraphics[width=\textwidth]{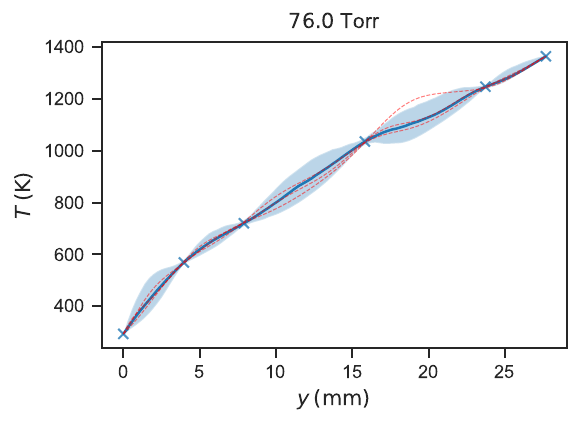}
        \caption{Posterior over $u$ at $76$ Torr.}        
    \end{subfigure}
    \begin{subfigure}[b]{0.475\textwidth}
        \centering
        \includegraphics[width=\textwidth]{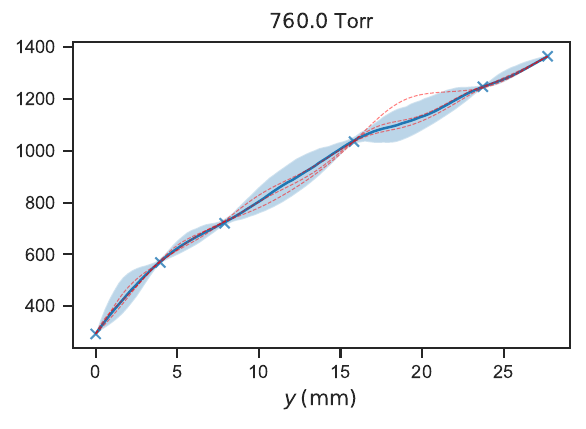}
        \caption{Posterior over $u$ at $760$ Torr.}        
    \end{subfigure}
    \caption{IFT temperature results across remaining experimental pressure values.}
    \label{fig:IFTtemp}
\end{figure}
\begin{figure}[h]
    \centering
    \begin{subfigure}[b]{0.475\textwidth}
        \centering
        \includegraphics[width=\textwidth]{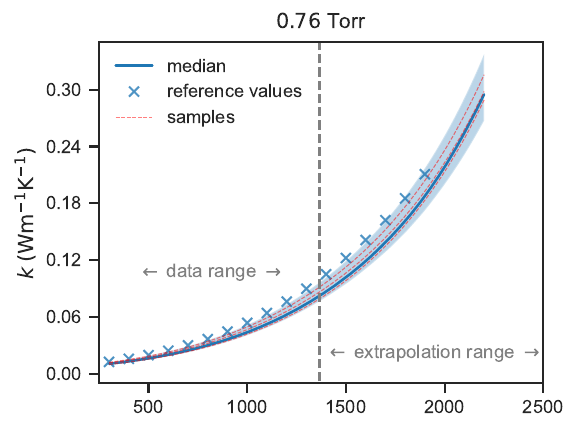}
        \caption{Posterior over $k$ at $0.76$ Torr.}        
    \end{subfigure}
    \begin{subfigure}[b]{0.475\textwidth}
        \centering
        \includegraphics[width=\textwidth]{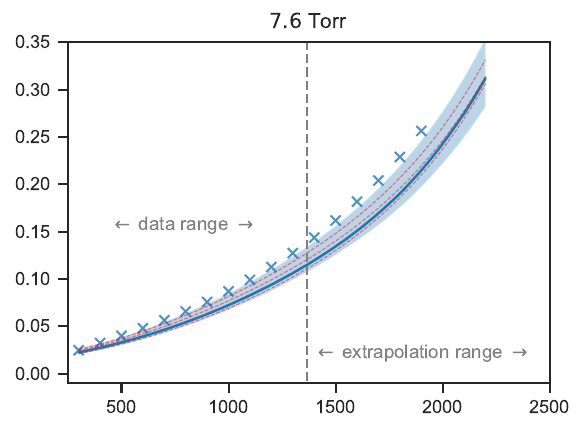}
        \caption{Posterior over $k$ at $7.6$ Torr.}        
    \end{subfigure}
    \hfill
    \begin{subfigure}[b]{0.475\textwidth}
        \centering
        \includegraphics[width=\textwidth]{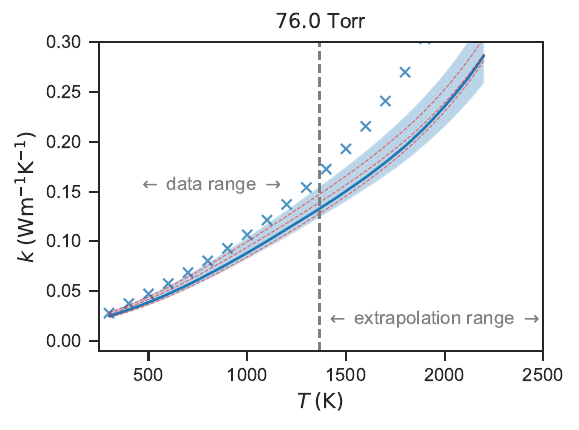}
        \caption{Posterior over $k$ at $76$ Torr.}        
    \end{subfigure}
    \begin{subfigure}[b]{0.475\textwidth}
        \centering
        \includegraphics[width=\textwidth]{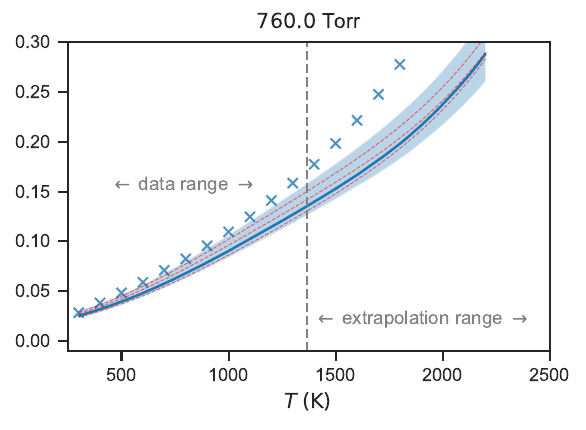}
        \caption{Posterior over $k$ at $760$ Torr.}        
    \end{subfigure}
    \caption{IFT thermal conductivity results across remaining experimental pressure values.}
    \label{fig:IFTk}
\end{figure}

For completion, we present the temperature and thermal conductivity IFT results across all experimental pressure values here.
The temperature results are shown in Fig.~\ref{fig:IFTtemp}.
We find that the method performs roughly the same across all experiments.
In Fig.~\ref{fig:IFTk}, we show the results for the thermal conductivity.
From the results, it appears to be the case that the thermal conductivity becomes more difficult to predict as the pressure approaches atmospheric pressure.
The reason the temperature reconstruction remains accurate regardless of the thermal conductivity fit is because the method sees temperature data.
Even though the thermal conductivity fit degrades at higher pressures, the entry simulations still remain robust, which could be attributed to the fact that the heat equation tends to be very forgiving.

We suspect that the inaccuracy at higher pressures is related to how the heat transfer behaves.
At low pressure levels, there is no gas conduction, meaning that the main source of heat transfer is radiation, especially at higher temperatures.
From our experiments, it seems most of the radiation effects can be captured by measuring up to $1367$ K.
However, as gas conduction begins to play a role, our model does not capture the nonlinear temperature-dependent density and thermal conductivity of gases and the pressure and temperature-dependent gas mean free path~\cite{kennard1938kinetic, marcussen1985thermal, daryabeigi2024thermal}.

One way to improve the fit could be to include an additive component of the thermal conductivity which comes from the gas conduction.
Instead, we tested how IFT solves the inverse problem if given experimental data with the hot end set to $1900$ K, taken from from~\cite{daryabeigi2024thermal}.
This provides additional latent information that changes how the thermal conductivity is learned.
\begin{figure}[h]
    \centering
    \begin{subfigure}[b]{0.475\textwidth}
        \centering
        \includegraphics[width=\textwidth]{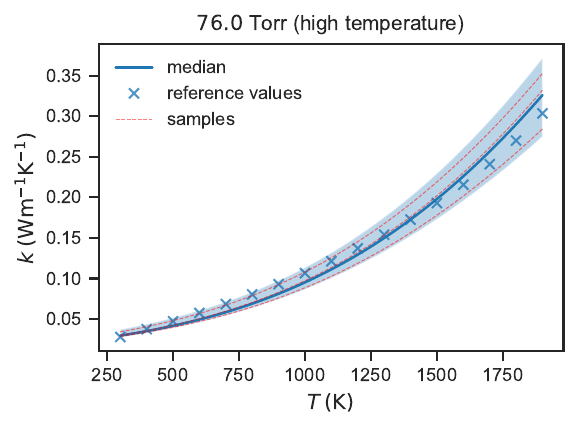}
        \caption{Posterior over $k$ at $76$ Torr.}        
    \end{subfigure}
    \begin{subfigure}[b]{0.475\textwidth}
        \centering
        \includegraphics[width=\textwidth]{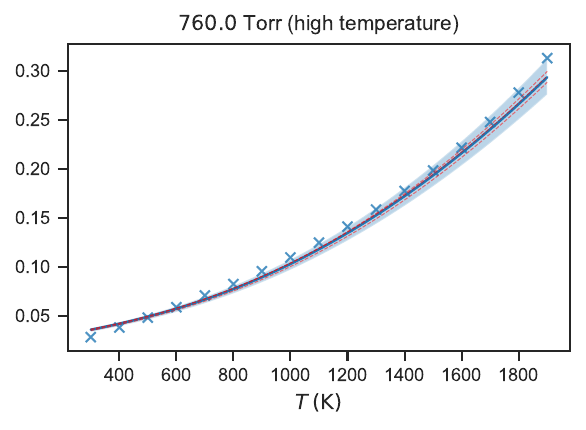}
        \caption{Posterior over $k$ at $760$ Torr.}        
    \end{subfigure}
    \caption{IFT thermal conductivity results at high pressures with a high temperature gradient (300 K on the cool end to 1900 K on the hot end).}
    \label{fig:highIFTk}
\end{figure}
We follow the same IFT setup as before to solve the inverse problem for the thermal conductivity with this larger temperature gradient.
The results for the highest two pressures, which the previous experiment could not accurately model, are shown in Fig.~\ref{fig:highIFTk}.
We find that the method performs better in this scenario, evidenced by the fact that the median qualitatively captures the overall trend and the reference thermal conductivity stays within the posterior distribution.
Note that this is not inherently a failing of IFT, rather it is a result of our modeling assumptions.
That is, we are ignoring the gas conduction in our treatment of the thermal conductivity parameterization.
Perhaps including this would improve the fit, although this would introduce additional complexity to the inverse problem.
And, as seen in our uncertainty propagation tasks, making this change is unnecessary for our purposes.

\section{Model validation}
\label{apdx:valid}

To validate the method, we solve the inverse problem on a synthetic dataset with a known temperature and thermal conductivity.
The synthetic data is generated from a digital twin of the steady state experiments performed at LaRC~\cite{daryabeigi2024thermal}.
We implement the digital twin and perform virtual experiments using the finite element method in the commercial software ABAQUS.
We perform a 2D simulation using linear continuum element (DC2D4 element type in ABAQUS).
When running the model, we take the reference OFI values as the ground truth, which is fed into the digital twin to generate corresponding temperature data.
We specify temperature boundary conditions on either end of the OFI material and perform a steady state heat transfer analysis.
Note that we only model the OFI layer from the experimental setup for this study, and not an entire TPS.
Once the analysis is complete, the temperature across the thickness of the OFI specimen is extracted and corrupted with a small amount of noise to generate synthetic datasets.

The data is generated at various static pressure levels, ranging from $0.001$ Torr to $760$ Torr, which mimics the real experimental setup.
We then use IFT to reconstruct the thermal conductivity of the OFI insulator from the synthetic temperature data, which is then compared to the reference values.
To solve the inverse problem with IFT, we apply the method exactly as is done with the raw experimental data (same hyperparameters, field parameterization, etc.).
\begin{figure}[h]
    \centering
    \begin{subfigure}[b]{0.475\textwidth}
        \centering
        \includegraphics[width=\textwidth]{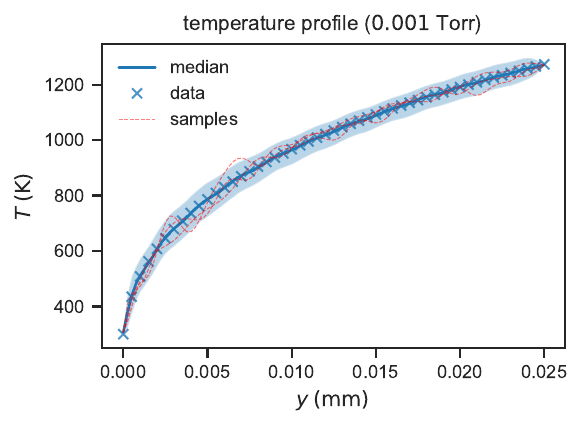}
        \caption{Posterior over the temperature.}        
    \end{subfigure}
    \begin{subfigure}[b]{0.475\textwidth}
        \centering
        \includegraphics[width=\textwidth]{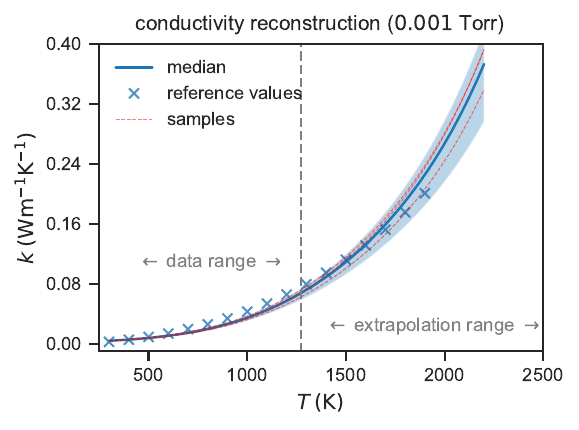}
        \caption{Posterior over the thermal conductivity.}
    \end{subfigure}
    \caption{IFT results at the lowest pressure, $0.001$ Torr, using the synthetic dataset.}
    \label{fig:lowpSynth}
\end{figure}

\begin{figure}[h]
    \centering
    \begin{subfigure}[b]{0.475\textwidth}
        \centering
        \includegraphics[width=\textwidth]{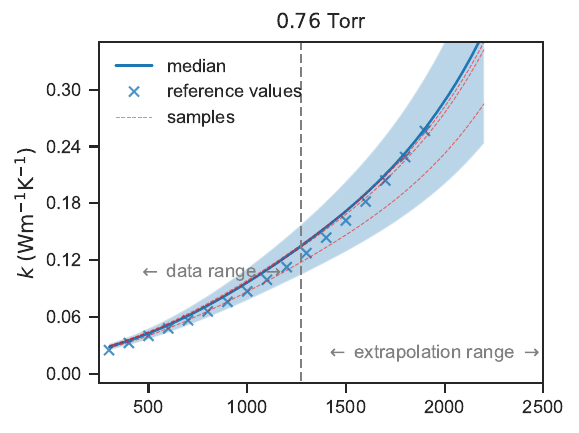}
        \caption{Posterior over $k$ at $0.76$ Torr.}        
    \end{subfigure}
    \begin{subfigure}[b]{0.475\textwidth}
        \centering
        \includegraphics[width=\textwidth]{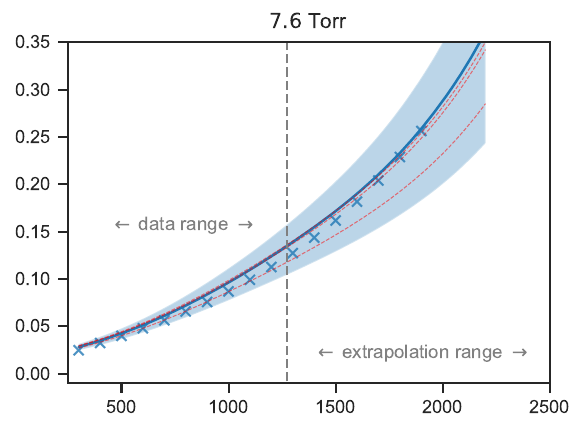}
        \caption{Posterior over $k$ at $7.6$ Torr.}        
    \end{subfigure}
    \hfill
    \begin{subfigure}[b]{0.475\textwidth}
        \centering
        \includegraphics[width=\textwidth]{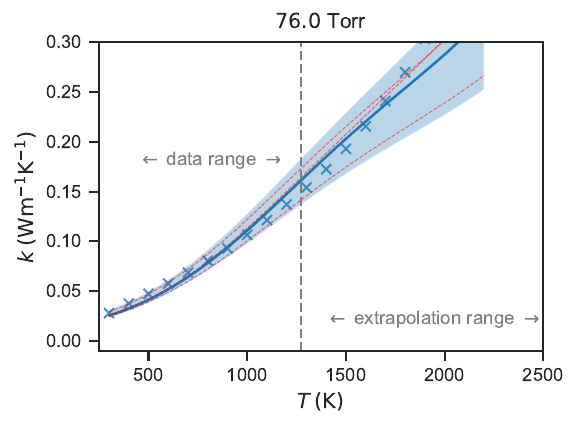}
        \caption{Posterior over $k$ at $76$ Torr.}        
    \end{subfigure}
    \begin{subfigure}[b]{0.475\textwidth}
        \centering
        \includegraphics[width=\textwidth]{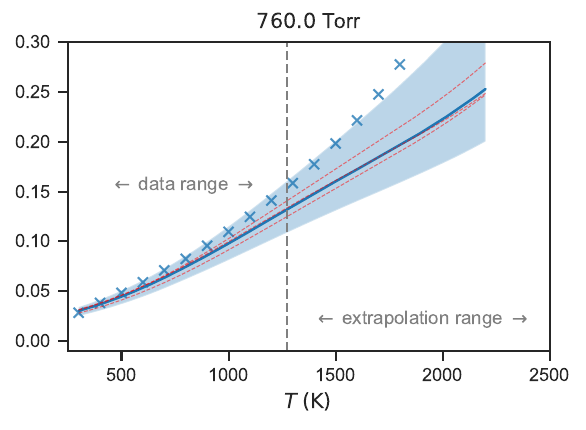}
        \caption{Posterior over $k$ at $760$ Torr.}        
    \end{subfigure}
    \caption{IFT thermal conductivity across remaining pressure levels with the synthetic data.}
    \label{fig:IFTk_synth}
\end{figure}

The resulting posterior obtained from IFT with the synthetic data at $0.001$ Torr is presented in Fig~\ref{fig:lowpSynth}.
We observe that the method accurately identifies the thermal conductivity, with a higher-fidelity reconstruction of the temperature profile when compared to the raw dataset.
This is to be expected, of course, as the use of a digital twin provides more measurements of the temperature than what was physically possible in the experiment.
The remaining thermal conductivity posteriors are shown in Fig.~\ref{fig:IFTk_synth}.
Just as with the experimental data, we see the accuracy of the reconstruction degrades at higher pressures.
However, there is slightly better performance with the synthetic data.
This is evidenced by the fact that the synthetic data largely remains within the posterior, which was not true with the experimental dataset.

\end{document}

%% file: intro.tex
\section{Introduction}
\label{sec:intro}

The sizing of thermal protection systems (TPS) of spacecrafts for atmosphere entry requires knowledge of flight trajectory and associated heating rates as well as thermal properties of the TPS.
The heat flux levels determine whether an ablative or passive TPS can be used to protect the spacecraft structure from the atmospheric entry heating.
Real world examples include ballistic entry of the Apollo spacecraft, which required ablative TPS~\cite{pavlosky1974apollo}, as opposed to the Space Shuttle lifting body glided entry, which enabled using reusable passive TPS~\cite{curry1993space}.
TPS consists of not only the insulation, but other components needed to attach to the underlying spacecraft structure to enable effective operation.
For example, the TPS on the windward side of the Space Shuttle mainly consisted of LI-900 insulation tiles, made of rigidized low-density high-purity silica fibers, with a reaction cured glass (RCG) coating applied to its outer mold line to increase its surface emittance.
A strain isolation pad (SIP) with thin layers of room-temperature vulcanizing silicone adhesive on either side of it was bonded to the tile and aluminum skin of the spacecraft to alleviate the thermal expansion mismatch between the ceramic tile and aluminum structure~\cite{curry1993space}.
The TPS in this case consisted of the tile, RCG coating, SIP, and the adhesive layers.
Thermal properties of the various components of the TPS as well as the underlying aluminum structure of the spacecraft were required for thermal analysis and sizing.
The main thermal design consideration was to size the TPS to ensure the underlying spacecraft aluminum structure would not exceed 175°C during entry and landing.
The more recent hypersonic inflatable aerodynamic decelerator (HIAD) concept utilized by NASA consists of an inflatable aeroshell to protect the payload during atmospheric entry~\cite{olds2013irve}, as opposed to the rigid aeroshell used on the Apollo.
The inflatable aeroshell consists of a flexible TPS covering an inflatable structure.
The aeroshell is packed and stowed, and is inflated and deployed prior to atmospheric entry.
The TPS sizing for HIAD requires thermal modeling of its various components: outer mold line fabric, insulation felts, and the inflatable structure.
The recent successful Low-Earth Orbit Flight Test of an Inflatable Decelerator (LOFTD) required sizing of the flexible TPS to ensure the inflatable structure would not exceed 400°C during entry~\cite{dinonno2024low}.

The accurate sizing of TPS to ensure the underlying structure does not exceed its temperature limit during atmospheric entry requires knowledge of the thermal properties of the TPS constituents over the temperature range experienced during flight.
Because insulation felts and fabrics are highly porous materials, their thermal properties are a function of not only temperature but the environmental pressure.
For passive insulations, thermal conductivity data over the static pressure range of 0.001 torr to 760 torr and temperature range of 300 K to 1900 K is typically required.
Various standard steady-state techniques~\cite{ASTM:C177-19,flynn1969radial,ASTM:C518} can be used for measuring the thermal conductivity of high-porosity thermal insulation samples.
The guarded-hot-plate technique~\cite{ASTM:C177-19} is the most accurate technique for testing of insulation samples but requires significant setup and test time to achieve steady-state conditions in order to yield accurate results.
Some guarded-hot-plate test setups can operate up to 1200 K~\cite{lee2019radiation}, but most of the test setups have upper temperature limits in the range of 600 K to 900 K~\cite{zarr2020nist,kratos_thermal_testing}.
The radial flow technique~\cite{flynn1969radial} can also be used for measuring thermal conductivity of insulation samples, but is currently limited to testing at atmospheric pressure for temperatures above 1000 K~\cite{kratos_thermal_testing}.
The heat flow meter technique~\cite{ASTM:C518} has a faster turnaround time compared to the guarded-hot-plate technique and has been used to measure thermal conductivity of various insulation samples up to 1900 K~\cite{daryabeigi2024thermal}.
Both guarded-hot-plate and heat flow meter techniques can provide thermal conductivity as a function of temperature and static pressure.
One transient technique, the three-point step heating technique, can provide accurate thermal diffusivity measurements of high-porosity insulation materials up to 1100 K at various static pressures~\cite{gembarovic2007method}.
Specific heat data can typically be obtained from the differential scanning calorimetry technique~\cite{ASTM:E1269}.  

Most of the current thermal conductivity measurement techniques applicable to high porosity insulations, with the exception of the heat flow meter setup at NASA Langley Research Center~\cite{daryabeigi2024thermal} (LaRC), can only provide data over a limited temperature range.
Given thermal properties over a limited temperature range, can one accurately estimate properties beyond the range of available data in order to accurately size a TPS?
Linear extrapolation of thermal conductivity data above the measured temperature range can potentially provide erroneous results since the thermal conductivity of insulation materials is highly nonlinear, especially at higher temperatures.
The heat transfer in a highly porous insulation mainly consists of radiation, gas conduction, and solid conduction modes of heat transfer~\cite{daryabeigi2011combined}.
The radiation component’s significance increases with increasing temperature and typically varies with temperature to the third power.

In this work, we approach the problem of estimating thermal conductivity of high-temperature insulation materials beyond their measured temperature range with information field theory (IFT).
IFT is a specialized technique for posing full-field Bayesian inverse problems~\cite{ensslin2009information}.
In this context, IFT works by treating the thermal conductivity as an unknown field (as a function of temperature and pressure) and uses Bayesian inference to derive an infinite-dimensional posterior distribution over the possible thermal conductivity configurations of the material which generated the experimental data.
We specifically rely on IFT with physics-informed priors~\cite{alberts2023physics}, which allows us to seamlessly integrate experimental data with the known physics of the problem, namely the heat equation.
Because IFT follows Bayesian principles, we also gain the ability to quantify the uncertainty in the thermal conductivity reconstruction.
This allows us to, for example, use the resulting posterior to propagate uncertainty through some important quantity of interest, such as the temperature of a spacecraft structure during an entry scenario.

We apply this technique to the Opacified Fibrous Insulation (OFI) felt whose properties had been reported over the temperature range of 300 K to 1900 K at various pressures between 0.001 torr and 760 torr in air~\cite{daryabeigi2024thermal}.
We compare the estimated values from our method to the published results, also referred to as the reference data.
But the ultimate comparison involves the sizing of a TPS for a specific flight trajectory using the reference thermal properties and the IFT estimated thermal properties in order to validate the method.
For this purpose, we simulate a HIAD flexible TPS configuration with the HIAD felts layers replaced by layers of OFI for two different flight entry simulations.
For these simulations, the OFI felt is encased between two layers of Hi-Nicalon silicon carbide outer fabric and a Kapton film gas barrier for the inflatable structure.
We size this flexible TPS using heating rates associated with the Mars Science Laboratory (MSL) mission~\cite{edquist2009aerothermodynamic, wright2014sizing}, and an International Space Station (ISS) return mission~\cite{walker2014preliminary}.
Note that a HIAD aeroshell was not designed for these missions; only the heating rates associated with the rigid aeroshells for these two missions were used to size the HIAD-like TPS.
The MSL mission had used a rigid aeroshell with phenolic impregnated carbon ablator (PICA) ablative TPS, while the ISS return study had used a rigid aeroshell with a multifunctional hot structure heat shield (MHSHS).

Our work is organized as follows.
In Sec.~\ref{sec:data}, we discuss how the different datasets, both experimental and synthetic, are collected.
The methodology is presented in Sec.~\ref{sec:ift}.
Here, we provide a brief background of infinite-dimensional Bayesian inverse problems in general and then describe the IFT methodology used to predict the thermal conductivity of the OFI.
We also show the results for various values of pressure.
To validate the method, we use the results obtained from IFT to simulate two different atmospheric entry scenarios in Sec.~\ref{sec:forward}.
We pose an uncertainty propagation task, where we use the thermal conductivity posterior samples to predict the temperature of a lander as it enters the atmosphere.
We design the insulation from real-world data, and compare the results to temperature profile generated with the ground truth thermal conductivity.

%% file: datageneration.tex
\section{Experimental setup and data curation}
\label{sec:data}

\begin{figure}[htbp]
    \centering
    \includegraphics[width=0.9\linewidth]{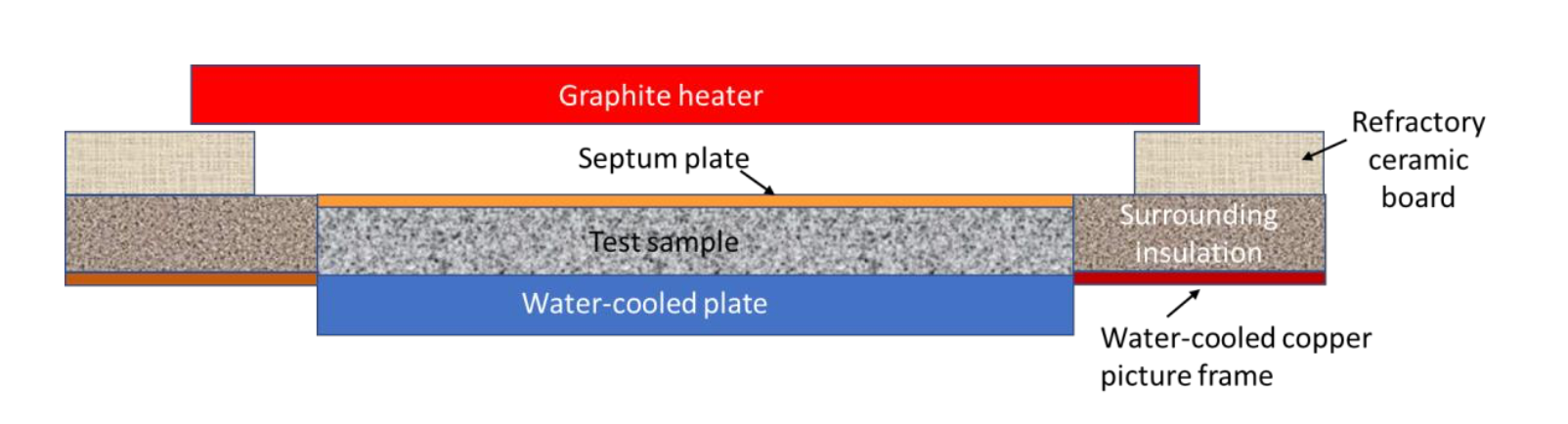}
    \caption{Illustration of experimental setup at NASA LaRC}
    \label{fig:exp_setup}
\end{figure}

We provide a short description of the heat flow meter experimental setup at LaRC.
An illustration of the setup is shown in Fig.~\ref{fig:exp_setup}.
The main components consist of a graphite heater, a graphite septum plate, the OFI test sample, and a water-cooled plate.
The OFI test sample is sandwiched between the graphite septum plate and the water-cooled plate.
When collecting the measurements, the septum plate temperature is maintained at discrete temperatures between $534$ K and $1853$ K, while the water-cooled plate is kept between $290 - 300$ K.
Test data are generated at various test pressures between $0.001$ torr and $760$ torr in nitrogen gas with the OFI sample subjected to temperature differences of $200$ K to $1600$ K across its nominal thickness of $2.77$ cm.
The measured temperatures of the septum plate and water-cooled plate, measured temperatures at various depths through the OFI test sample thickness, and measured heat flux on the water-cooled plate at steady-state conditions are used to determine the thermal conductivity of the test sample as a function of temperature for the test static pressure.
More details on the test procedure and data processing can be found in the report~\cite{daryabeigi2024thermal}.
For the thermal conductivity, we also take the published results in the report as reference values, which were identified using regression with cubic polynomials.

To investigate the use of IFT in estimating the thermal conductivity of the fibrous insulation we compare the resulting posterior distributions with the reference values.
Our goal is to evaluate the effectiveness of IFT in extrapolating the thermal conductivity predictions with uncertainty in regions beyond experimental capabilities.
This is in effort to simulate entry scenarios, where the temperature on the outer mold line of the TPS is expected to reach temperatures close to 2000 K.
For this reason, we take as experimental data the case where the temperature of the hot end is set to $1367$ K, which allows us to set aside higher temperature reference values of thermal conductivity for comparison.
Using the temperature data, IFT reconstructs the thermal conductivity of the OFI insulator with quantified uncertainty.
This works by obtaining a posterior distribution over the possible thermal conductivity fields of OFI.
The posterior is then used to perform an uncertainty propagation task, where we look to predict the temperature across the body of a lander in atmospheric entry scenarios.

%% file: ift.tex
\section{Information field theory framework}
\label{sec:ift}
The task at hand is to reconstruct the thermal conductivity of the OFI at high temperatures (above $1367$ K) relying only on low temperature measurements, which is an inverse problem.
Mathematically, this inverse problem appears in the usual form as
$$
y = R(k) + \xi,
$$
where $y$ is a collection of temperature measurements across the body, $R$ is a map which takes the thermal conductivity $k$ as input and outputs predictions of the temperature, and $\xi$ is additive noise, which we will take to be Gaussian.
Solving this inverse problem involves inverting the map between $k$ and $y$, i.e., identify the thermal conductivity which generated the data.
Inverse problems are most often ill-posed, meaning that they oftentimes do not have a unique solution.
In our case, we do not measure $k$ directly, only a byproduct (the temperature).
That is, the measurement operator $R$ is described by the solution of the nonlinear heat equation at the estimate of $k$ and then evaluated at the measurement locations.
This is an inverse problem which is famously ill-posed~\cite{ghattas2021learning}.
We also look to extrapolate the model predictions beyond regions where data is collected, adding further complications.

Due to the ill-posed nature of the inverse problem at hand, we turn to the Bayesian approach, which formulates the problem in the language of probability~\cite{kaipio2006statistical, stuart2010inverse}.
As we will see, what results is a probability distribution over the possible $k$'s.
This represents the degree of uncertainty about the ground truth thermal conductivity.
There is a large collection of techniques for solving Bayesian inverse problems, e.g., Markov chain Monte Carlo (MCMC) in function space~\cite{cotter2013mcmc} or the method of adjoints~\cite{ghattas2021learning}.
However, these methods are computationally prohibitive without considerable effort, as evaluating the input to output map $R$ requires solving the partial differential equation (PDE) which governs the system.
This means that we must solve the forward problem for every sample from the posterior which would require calling our finite element (FE) solver thousands of times.
Other techniques avoid doing so by working with the \emph{residuals} of the PDEs with Gaussian process (GP) regression~\cite{williams1995gaussian,chen2021solving}, Bayesian physics-informed neural networks~\cite{yang2021b}, and more.
However, the GP approach provides only a maximum a posteriori estimate, meaning there is no uncertainty quantification about the solution to the inverse problem, and there is evidence that Bayesian-PINNs do not extrapolate well to regions without data~\cite{hao2025neural}.

Our approach relies on IFT with physics-informed priors~\cite{alberts2023physics}.
IFT is a scientific machine learning framework for Bayesian inverse problems formulated in the infinite-dimensional setting, first introduced for cosmological imaging applications~\cite{ensslin2009information}.
Like the standard methods for Bayesian inverse problems, IFT works by defining a prior probability measure over the space of possible fields.
Compared to other Bayesian approaches, the IFT approach has a number of advantages.
In most cases, a GP prior is chosen, see for example~\cite{cotter2010approximation, bui2013computational, chen2021solving, bai2024gaussian}.
This requires selection of a covariance kernel, which is often done in an ad hoc manner and places added assumptions on the field.
The physics are also pushed to the measurement model, creating undesirable effects in the case of model-form error, scalability, etc.~\cite{alberts2023physics}.
Unlike the standard paradigm, the IFT prior is defined directly from the physics.
This also means that the forward model does not appear in the likelihood, which allows IFT to scale well.
Further, the solution to the forward map i.e., the temperature field, is also learned as a part of the inverse problem and inferred from data.
This is perhaps the main difference between IFT and other methods.
Finally, methods such as Bayesian-PINNs start from a `discretization first' viewpoint, which can bias the prior.
In IFT, we start directly from the physics, and derive the entire methodology in the infinite-dimensional setting.
The specific discretization scheme when implementing code is then up to the user, allowing our method to generalize well.

\subsection{Construction of the IFT posterior}

Since the ultimate goal is to reconstruct the thermal conductivity of OFI from data, we need to derive the joint posterior between the temperature and the thermal conductivity.
The method begins by deriving a physics-informed prior over the space of possible temperature fields, conditional on the thermal conductivity.
This prior is built from the PDE.
Under our experimental apparatus, the heat transfer occurs only through the layers of the material.
The specimen also reaches steady-state before data is recorded.
Hence, we take as our model the one-dimensional steady-state heat equation
\begin{align}
    \label{eqn:heateqn}
    \frac{\partial}{\partial y}\left ( k(u) \frac{\partial u}{\partial y}\right) &= 0, \nonumber \\
    u(y_0) &= u_0, \nonumber \\
    u(y_{\ell})&=u_{\ell},
\end{align}
where $u$ is the temperature across the body and the thermal conductivity, $k$, is known to vary with temperature.
The boundary conditions for $u$ are known, as they are given by the temperatures of the hot and cold plates.

Using techniques from IFT, we build a physics-informed prior for $u$ conditional on $k$ from the heat equation.
To do so, we first turn eq.~(\ref{eqn:heateqn}) into an optimization problem via Dirichlet's principle
\begin{equation}
    \label{eqn:ISR}
    \left\{
    \begin{split}
        & \min_{u} \quad  E(u;k) \coloneqq \int_{y_0}^{y_{\ell}} \frac{\kappa (u)}{2}\left|\frac{\partial u}{\partial y}\right|^2dy \\
       & \textrm{subject to} \quad  u(y_0) = u_0, \quad u(y_{\ell}) = u_{\ell}
    \end{split}
    \right.
\end{equation}
This variational form provides an optimization problem which provides the solutions to eq.~(\ref{eqn:heateqn}), and it serves as the starting point to define loss functions in deep Ritz~\cite{yu2018deep} and scientific machine learning~\cite{karniadakis2021physics}.
That is, given $k$, a field $u$ which solves eq.~(\ref{eqn:ISR}) is also the unique solution to the heat equation.

Taking the variational form, we define the physics-informed prior for $u$ using the following Gibbs measure
\begin{equation}
    \label{eqn:prior}
    p(u | k) = \frac{1}{\mathcal{Z}(k)} \exp\left\{-\beta E(u;k)\right\},
\end{equation}
where $\beta \geq 0$ is a tunable hyperparameter.
The normalization constant of eq.~(\ref{eqn:prior}) is called the partition function and is defined via a path integral $\mathcal{Z}(k) = \int \mathcal{D}u\: \exp\left\{-\beta E(u;k)\right\}$.
Path integrals are an application of functional integration~\cite{cartier2006functional} for defining probability measures over function spaces~\cite{feynman2010quantum, albeverio1976mathematical}, and are commonly encountered in quantum~\cite{glimm2012quantum} and statistical field theories~\cite{parisi1988statistical}.
Since eq.~(\ref{eqn:prior}) is conditional on $k$, the partition function will depend on the current value of $k$.
This presents a major issue, as in almost all cases path integrals cannot be directly computed.
This will become apparent later.
Fortunately, there are specialized sampling algorithms designed for IFT which avoid the need to characterize the partition function.

The form of physics-informed prior given by eq.~(\ref{eqn:prior}) is chosen so that higher probability density is assigned to temperature fields which satisfy the physics, directly encoding our prior knowledge that $u$ should obey the heat equation.
This is because a field $u$ which minimizes eq.~(\ref{eqn:ISR}) maximizes the prior density.
The hyperparameter $\beta$ appearing in the physics-informed prior is referred to as the model trust in IFT literature.
The variance of samples from the physics-informed prior is controlled by the magnitude of $\beta$, and it can be viewed as a measurement of the agreement between the chosen physical model and the ground truth physics.
The limiting behavior of $\beta$ is studied in great detail in~\cite{alberts2024well}, where a guideline for picking the value of $\beta$ in numerical experiments is posed.

Next, we describe the prior for the thermal conductivity.
Since $k$ is unobserved, it is crucial to encode as much information as possible into the prior for accurate reconstruction.
In this specific application, we know $k$ should be an increasing function of temperature.
Hence, we sample in the $\log$-space so that $k$ remains positive during training.
We also know from experimental data roughly the order of magnitude and length scale of $k$.
In practice, one could first identify a deterministic estimate of $k$ to gain more insight before assigning a prior and applying IFT.
We choose a transformed GP with the squared exponential kernel for $k$
\begin{equation}
    \label{eqn:kprior}
    \log p(k) = \mathcal{GP}(k|c_1 + c_2u, s),
\end{equation}
where we have chosen a linear mean, which signifies that we expect $k$ to grow exponentially with $u$, as the prior is defined in the $\log$-space.
The squared exponential kernel has the form
$$
s(u,u') = \exp\left(-\frac{(u-u')^2}{2 \ell^2}\right),
$$
where $\ell$ is the lengthscale parameter.
The hyperparameters of the GP prior are $(c_1, c_2, \ell)$, which we optimize through a maximum likelihood estimate before sampling from the full posterior.

The likelihood is a model of the data-generating process, which we take to be a simple zero-mean Gaussian measurement model.
The temperature data follows what was outlined in Sec.~\ref{sec:data}.
That is, the temperature of the material is collected at points $y_i$, $i = 1,\dots,m$, with i.i.d. Gaussian noise with a standard deviation of $\xi = 2.5\:\mathrm{K}$.
Let $\mathbf{d}$ denote the vector of individual temperature measurements collected, $\mathbf{d} = (d_1,\dots, d_m)$.
The likelihood model then decomposes into a product of Gaussians:
\begin{equation}
    \label{eqn:likelihood}
    p(\mathbf{d}|u) = \prod_{i=1}^m\mathcal{N}(d_i|u(y_i),\xi^2). 
\end{equation}

Finally, application of Bayes's rule between the priors eqs.~(\ref{eqn:kprior}) and~(\ref{eqn:prior}) and the likelihood~(\ref{eqn:likelihood}) reveals the joint posterior
\begin{equation}
    \label{eqn:posterior}
    p(u, k | \mathbf{d}) = \frac{1}{p(\mathbf{d})}p(\mathbf{d}|u)p(u|k)p(k),
\end{equation}
where $p(\mathbf{d})$ is the normalization constant.
The joint posterior quantifies our state of knowledge about $u$ and $k$ after we have observed the temperature data.
It is important to note here that the normalization constant of the prior, $\mathcal{Z}(k)$, is hidden within the posterior, and cannot be ignored.
This is a consequence of the fact that we have chosen to condition the physics-informed prior $p(u|k)$ on $k$, therefore its normalization constant will vary at each value $k$, which is inferred.
For sampling, characterizing $\mathcal{Z}(k)$ would require evaluation of a path integral, which to the best of our knowledge, is not possible in this case.
Instead, we use the sampling scheme derived in~\cite{alberts2023physics}, which avoids this problem.

\subsection{Numerical sampling scheme}
To sample from the joint posterior in eq.~(\ref{eqn:posterior}), we implement the nested stochastic gradient Langevin dynamics (NSGLD) scheme derived in~\cite{alberts2023physics}.
The nested scheme uses two loops of SGLD~\cite{welling2011bayesian} wrapped on top of each other.
In the inner loop, we generate new samples of $u$ with the most recent sample of $k$.
On top of this, we generate a new sample of $k$ using the inner loop samples of $u$.
In each loop, the method only requires a noisy estimate of the gradient of the individual marginal posteriors, which allows the method to scale well.
We summarize NSGLD below.

For a generic probability distribution $p(\theta)$, SGLD generates a new sample $\theta_{t+1}$ according to
\begin{equation}
    \label{eqn:SGLD}
    \theta_{t+1} = \theta_t + \frac{1}{2}\varepsilon_t\nabla_{\theta}\log p(\theta_t) + \eta_t,
\end{equation}
where $\eta_t \sim \mathcal{N}(0,\varepsilon_t)$ is injected noise, and the learning rate $\varepsilon_t$ satisfies the Robbins-Monro conditions~\cite{robbins1951stochastic}
$$
\sum_{t=1}^{\infty}\varepsilon_t=\infty, \quad \sum_{t=1}^{\infty}\varepsilon_t^2<\infty.
$$
The power of SGLD is that unlike most Markov chain Monte Carlo algorithms, SGLD  generates samples from the underlying distribution using an unbiased estimate of $\log p(\theta_t)$.
In terms of a likelihood, one may recognize this as allowing for minibatching.
We exploit this fact to transform the joint posterior eq.~(\ref{eqn:posterior}) into a suitable form to generate samples following eq.~(\ref{eqn:SGLD}) by taking Monte Carlo estimates of the various integrals that appear.

The first step is to separate the joint and write the marginal posterior over the thermal conductivity $k$,
\begin{equation}
    \label{eqn:margk}
    p(k | \mathbf{d}) = \int \mathcal{D}u\: p(u,k|\mathbf{d}).
\end{equation}
Similarly, we can write for the temperature $u$
\begin{equation}
    \label{eqn:margu}
    p(u|\mathbf{d},k) \propto p(\mathbf{d}|u)p(u|k),
\end{equation}
where we may drop $\mathcal{Z}(k)$ since it is a constant in eq.~(\ref{eqn:margu}).

We now parameterize the fields.
Under IFT, the specific choice is up to the user, e.g., one could pick neural networks, finite element nodes, basis functions, etc.
For the temperature, we choose a truncated Fourier series
$$
\hat{u}_{\theta}(y) = y(1-t)\left(\theta_1 + \sum_{j=1}^N\left\{\theta_{j+1}\cos(2\pi jy) + \theta_{j+N+1}\sin(2\pi j y)\right\}\right) + u_{\ell}y + (1-y)u_0,
$$
which is modified so that the boundary conditions are a priori satisfied.
Recall that for the thermal conductivity, we have chosen a GP prior.
Then, a natural field parameterization for $k$ is the truncated Karhunen-Lo\`eve expansion (KLE) of the GP~\cite{williams2006gaussian}.
This is a spectral decomposition which expresses $k$ in terms of the eigenvalues and eigenfunctions of the covariance kernel.

Letting $S$ be the integral operator with kernel $s$, the KLE states that a draw from the GP $f \sim \mathcal{GP}(m,s)$ can be generated according to
$$
f(\cdot) = m(\cdot) + \sum_{n \in \mathbb{N}}z_n\lambda_n^{1/2}\phi_n(\cdot),
$$
where $z_n\overset{\mathrm{i.i.d.}}{\sim}\mathcal{N}(0,1)$, and $\lambda_n$ and $\phi_n$ are the eigenvalues and corresponding eigenvectors of $S$.
That is, for each $n$, they satisfy
\begin{equation*}
    \label{eqn:Fred}
    \int s(\cdot,x')\phi_n(x')p(x')dx' = \lambda_n\phi_n,
\end{equation*}
which is a Fredholm integral equation of the first kind, weighted by some probability density $p(x)$.
For many kernels, the eigen-pairs are not analytically available, and some approximation to eq.~(\ref{eqn:Fred}) is used instead, such as the Nystr\"om method~\cite{quinonero2005unifying}.

For the thermal conductivity, we select the squared exponential kernel $s(u,u') = \exp\left(-(u-u')^2/2 \ell^2\right)$, which allows us to include the expected lengthscale $\ell$ as an inductive bias.
In actuality, we model $\log k$ as a GP with the squared exponential kernel, so that the thermal conductivity samples remain positive.
We also fit a linear mean function $m(u) = a_1 + a_2u$ for $\log k$, which translates to an exponential mean for $k$.
Putting everything together, a draw for $k$ can be generated from
\begin{equation}
    \label{eqn:kprior}
    k(u) = \exp\left(a_1 + a_2u + \sum_{n \in \mathbb{N}}z_n\lambda_n^{1/2}\phi_n(u)\right),
\end{equation}
where $z_n\overset{\mathrm{i.i.d.}}{\sim}\mathcal{N}(0,1)$.
We truncate the expansion to a finite number of terms $M$ for sampling.
The eigenvalues and eigenfunctions are taken from a transformation of the Hermite polynomials, the details of which are left to Appendix~\ref{apdx:A}.
We also assign a standard normal prior to both $a_1$ and $a_2$.

The joint IFT posterior can be characterized by sampling the field parameters $\theta \in \mathbb{R}^{2N+1}$ for the temperature and $a_1,a_2\in \mathbb{R}$ together with $z\in\mathbb{R}^M$ for the thermal conductivity.
To simplify the notation, we will represent the thermal conductivity parameters as the grouping $\lambda = (a_1,a_2,z_1,\dots,z_M)$.
Returning to the marginal posteriors, we replace the fields with their respective parameterizations and obtain
\begin{align}
    p(\lambda|\mathbf{d}) &= \int p(\theta,\lambda|\mathbf{d)}d\theta \\
    p(\theta|\mathbf{d},\lambda) &\propto p(\mathbf{d}|\theta)p(\theta|\lambda).
\end{align}

In~\cite{alberts2023physics}, it is shown that 
\begin{equation}
    \label{eqn:gradient}
    \nabla_{\lambda} \log p(\lambda|\mathbf{d}) = \mathbb{E}_{\theta\sim p(\theta|\lambda)}\left[\nabla_\lambda\beta E(u;k)\right] - \mathbb{E}_{\theta\sim p(\theta|\mathbf{d},\lambda)}\left[\nabla_\lambda\beta E(u;k)\right] - \nabla_{\lambda}\log p(\lambda),
\end{equation}
which provides a gradient step to sample the posterior over the thermal conductivity $p(\lambda|\mathbf{d})$ with SGLD, see eq.~(\ref{eqn:SGLD}).
For NSGLD, we build an unbiased estimate of eq.~(\ref{eqn:gradient}) via sampling averages of the expectations over $\theta$.
To see this, we start with the posterior over temperature,  $p(\theta|\mathbf{d},\lambda)$, where we will construct an unbiased estimate of $\log p(\theta|\mathbf{d},\lambda)$.
Explicitly, we have
\begin{equation}
    \label{eqn:hamil}
    \log  p(\theta|\mathbf{d},\lambda) = -\frac{1}{2\xi^2} \sum_{i=1}^m\left(\hat{u}_{\theta}(y_i) - d_i\right)^2 -\beta \int_{0}^{1}\frac{k(\hat{u}_{\theta};\lambda)}{2}\left|\frac{\partial \hat{u}_{\theta}}{\partial y}\right|^2dy + \mathrm{const.},
\end{equation}
where the additive constant contains both the normalization constant of the likelihood and the partition function $\mathcal{Z}(\lambda)$.
Observe here that since $k$ is fixed in this step, $\mathcal{Z}(k)$ can be treated as a constant for the purposes of generating a sample from $p(\theta|\mathbf{d},\lambda)$ and dropped.
The likelihood term can be minibatched as usual, and for the physics, we take an importance sampling estimate of the integral.
That is, we take
$$
E(\hat{u}_\theta;k(\hat{u}_{\theta},\lambda)) \approx \frac{1}{b}\sum_{i=1}^b \frac{k(\hat{u}_{\theta};\lambda)}{2}\left|\frac{\hat{u}_{\theta}}{\partial y}\right|^2\Big|_{y=Y_i},
$$
with each $Y_i$, $i=1,\dots,b$, sampled uniformly in the domain.
Hence, we now have a way to generate samples of the posterior over the temperature through SGLD.
Once a collection of temperature samples are generated with $k$ fixed to its current sample, we build a sampling average approximation of the expectations appearing in eq.~(\ref{eqn:gradient}) to push $k$ forward.
Again, we replace $E(\hat{u}_\theta;k(\hat{u}_{\theta},\lambda))$ with the Monte Carlo estimate using the new sample of $k$.
Note that we also need samples of the physics-informed prior for the first expectation appearing in eq.~(\ref{eqn:gradient}), which can simply be generated in the same manner as the posterior by removing the likelihood term in eq.~(\ref{eqn:hamil}).
We then iterate back and forth between the temperature and conductivity marginal posterior SGLD steps to complete the NSGLD algorithm.

\subsection{Results}

When solving the inverse problem, we first nondimensionalize the problem and scale the temperature data so that it falls between $0$ and $1$.
We then rescale back into the original units when reporting the results.
Before sampling from the posterior we initialize the temperature field parameters to the maximum a posteriori estimate
$$
\theta_0 = \arg\max \log p(\theta|\mathbf{d},\lambda_0).
$$
This is done in order to speed up the rate at which SGLD approaches regions of high probability and to improve numerical stability.
To identify the MAP estimate, we implement the Adam scheme~\cite{kingma2014adam} built into the Optax library~\cite{deepmind2020jax}.
The conductivity is initialized by taking a random sample $\lambda_0$ from the prior described in eq.~(\ref{eqn:kprior}).

We now discuss the hyperparameters used when sampling the posterior.
Across all pressure values for which we generate the posterior, we keep all the hyperparmeters the same to tune as little as possible.
For the thermal conductivity prior, we keep the first $M=25$ terms of the expansion, which we find is sufficient to capture the overall trend of the thermal conductivity.
For the temperature expansion, we keep $N= 20$ terms and we set $\beta = 1000$, which we found to empirically give the best temperature reconstruction.
For the inner-loop sampling learning rate (which generates samples of $\theta$), we select
$$
\varepsilon_t = \frac{\alpha_1}{(t+\alpha_2)^{\alpha_3}},
$$
which is a standard choice throughout the literature.
We set $\alpha_2 = 0.01$ and $\alpha_3 = 0.51$ which ensures that $\varepsilon_t$ satisfies the Robbins-Monro conditions.
We then set $\alpha_1 = 0.1/\beta$, as in applications of IFT, letting the learning rate be inversely proportional to $\beta$ seems to be the correct choice~\cite{alberts2023physics}.
For the outer loop, which generates samples of $\lambda$, we pick the same configuration.

\begin{figure}[htbp]
    \centering
    \begin{subfigure}[b]{0.475\textwidth}
        \centering
        \includegraphics[width=\textwidth]{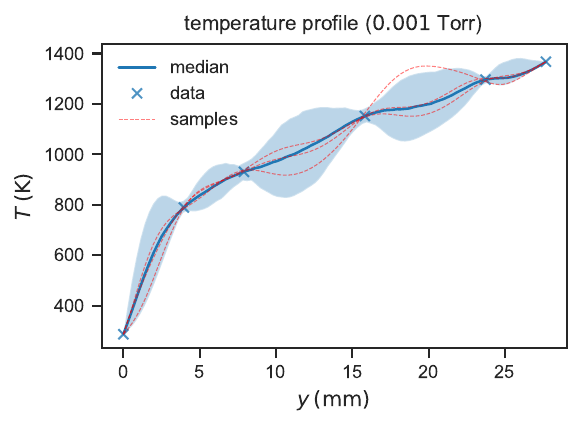}
        \subcaption{Posterior over $u$ at $0.001$ Torr.}
    \end{subfigure}
    \hfill
    \begin{subfigure}[b]{0.475\textwidth}
        \centering
        \includegraphics[width=\textwidth]{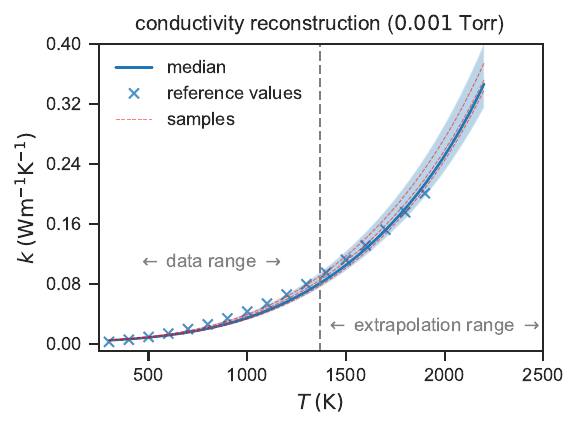}
        \subcaption{Posterior over $k$ at $0.001$ Torr.}
    \end{subfigure}
    \caption{IFT results at $0.001$ Torr. For each state, we record the resulting posterior median field, the $95\%$ central credible interval, and a few samples from the posterior.}
    \label{fig:001posterior}
\end{figure}

Across all experimental pressure values, we generate $100,000$ samples of the joint posterior and burn the first $10\%$.
The results at the lowest pressure, $0.001$ Torr, are shown in Fig.~\ref{fig:001posterior}, with the remaining results summarized in Appendix~\ref{apdx:B}.
We also validate the method on synthetic datasets derived from a digital twin of the experimental setup, which is discussed in Appendix~\ref{apdx:valid}. 
We record the posterior median and the central $95\%$ credible interval, along with a few samples of each field.
Recall that when generating the posterior, we only use temperature data up to $1367$ K and extrapolate the conductivity predictions to temperatures up to $2000$ K, which is required for the entry simulations.
This is done by numerically characterizing the posterior predictive distribution from the samples.
We find that the method generates good agreement with the ground truth.
This is due to the fact that the approach is physics-based, so the predictions remain physically reasonable beyond regions where data is available.
Also of note is the fact that the uncertainty grows as the temperature moves further away from the experimental region.

The major advantage gained by using the IFT approach, as opposed to a standard method, e.g., MCMC with Gaussian processes, in this application is that we have removed the need for a forward solver when solving the inverse problem.
In the traditional approach, one would sample the Gaussian process prior for $k$, which is then passed to the forward solver to evaluate the likelihood.
Although our FE solver is not too cumbersome, as it takes less than 5 minutes to generate the temperature profile for a given thermal conductivity, such an approach is unreasonable for this application without considerable effort/resources.
To get a clear picture of the posterior distribution, MCMC typically requires on the order of $10,000$ samples.
With IFT, we remove the forward solver completely by working with the PDE residuals instead.
This also allows for easy parallelization, something which we take advantage of, as each PDE residual can be computed in parallel.

%% file: TPS.tex
\section{Uncertainty propagation through a TPS in entry scenarios}
\label{sec:forward}

To further demonstrate the usefulness of the method, we present an application where the resulting posterior distribution over the OFI thermal conductivity is used to evaluate a TPS design.
We cover two different entry scenarios, one for Mars and one for Earth, using heat flux and pressure data taken from~\cite{walker2015multifunctional}.
By propagating the posterior uncertainty of the thermal conductivity through a digital twin of a lander, we can predict, with uncertainty, the temperature through the body of the TPS as it enters the atmosphere.

\subsection{Finite element simulation details}
\begin{wrapfigure}{r}{0.5\textwidth}
    \centering
    \includegraphics[width=0.8\linewidth]{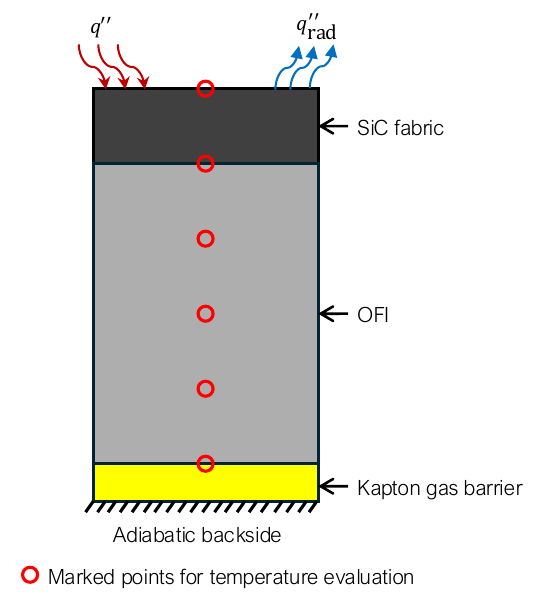}
    \caption{TPS setup used for entry simulations.}
    \label{fig:TPS_system}
\end{wrapfigure}

In both scenarios, we use a similar TPS, and perform a 1D transient heat transfer analysis.
We restrict ourselves to the 1D case, which is common practice in TPS analysis.
However, we have made our code open source and implemented the simulation so that any geometry and loading conditions may be used.
The focus is on estimating temperature profiles at the marked points in the TPS, shown in Fig.~\ref{fig:TPS_system}.
We compare the temperature profile the TPS using the reference thermal conductivity values and the IFT estimated thermal conductivities for the OFI.
In addition to this, since IFT provides a Bayesian posterior distribution, we propagate the uncertainty about the thermal conductivity reconstruction through the digital twin.
This allows us to quantify the uncertainty in the temperature throughout the body of the TPS during the entry simulation.

We begin by implementing a custom FORTRAN subroutine, which we call UMATHT, in ABAQUS to model the temperature and pressure dependent thermal conductivity. 
All the material files and subroutines are available at \url{https://github.com/PredictiveScienceLab/IFT-Thermal-Conductivity-Experiments}.
Both the SiC and OFI have temperature and pressure dependent thermal conductivity.
To model this behavior, we first populate a material file with a grid of values representing the thermal conductivity evaluated at various temperatures and pressures.
We use the stagnation pressure as a field variable applied on all the nodes of the FE model.
The stagnation pressure is a function of time and is modeled as a discrete amplitude curve.
The UMATHT subroutine evaluates the thermal conductivity of both the SiC and the OFI at the current time increment with a bilinear interpolation scheme using the grid of data supplied to the material file.
For the Kapton film, the subroutine is not invoked, as the thermal conductivity is a simple function of temperature.

We now discuss the material properties used for the simulations.
The specific heat (in $\text{J}\: \text{kg}^{-1}\text{K}^{-1}$) for both the SiC and OFI is modeled as a function of temperature T (in K) given as
\begin{equation}
    C_p = c_1 \left(1 - \frac{1}{\exp{\left(c_2 T\right)}}\right).
\end{equation}
For SiC fabric, the value for $c_1$ and $c_2$ are 1315 and 2.57e-3, respectively.
For OFI,  the value for $c_1$ and $c_2$ are 1302.4 and 2.73e-3, respectively.
The density of the SiC fabric, OFI, and the Kapton gas barrier is 799 kg m$^{-3}$, 99.4 kg m$^{-3}$, and 932  kg m$^{-3}$, respectively.
Even though the emissivity of SiC is temperature dependent, FE calculations were made using a constant SiC fabric emittance of 0.8.

\subsection{Mars entry simulations}

\begin{figure}[h]
    \centering
    \begin{subfigure}[b]{0.33\textwidth}
        \centering
        \includegraphics[width = \textwidth]{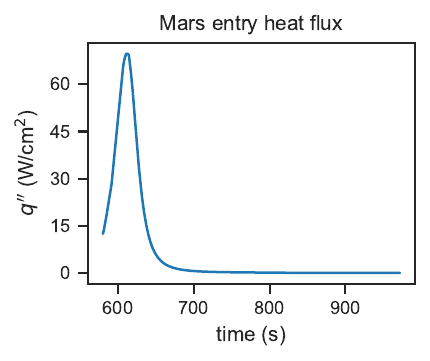}
        \caption{Mars heating profile}
    \end{subfigure}
    \begin{subfigure}[b]{0.33\textwidth}
        \centering
        \includegraphics[width = \textwidth]{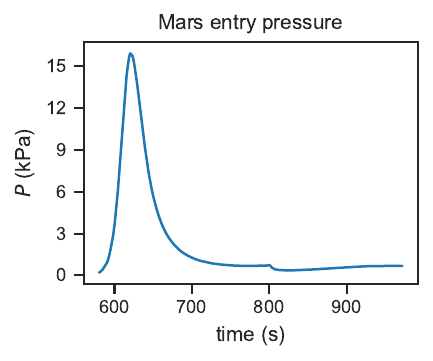}
        \caption{Mars stagnation pressure data}
    \end{subfigure}
    \caption{MSL cold wall heat flux and stagnation used for sizing TPS for Mars entry.}
    \label{fig:marsdata}
\end{figure}

\begin{figure}[htbp]
    \centering
    \begin{subfigure}[b]{0.3\textwidth}
        \centering
        \includegraphics[width=\textwidth]{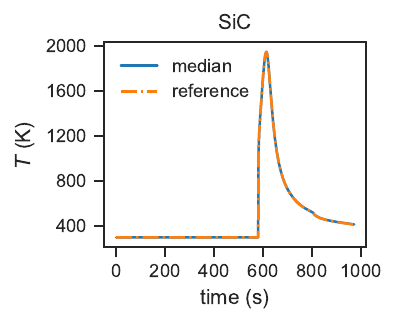}
    \end{subfigure}
    \begin{subfigure}[b]{0.3\textwidth}
        \centering
        \includegraphics[width=\textwidth]{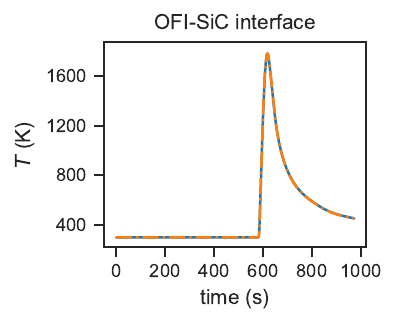}
    \end{subfigure}
    \begin{subfigure}[b]{0.3\textwidth}
        \centering
        \includegraphics[width=\textwidth]{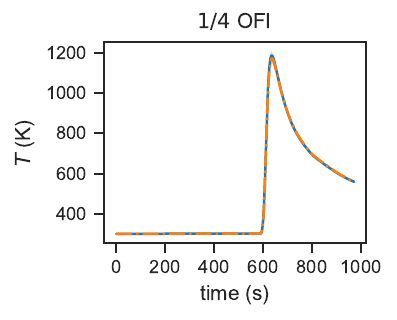}
    \end{subfigure}
    \hfill
    \begin{subfigure}[b]{0.3\textwidth}
        \centering
        \includegraphics[width=\textwidth]{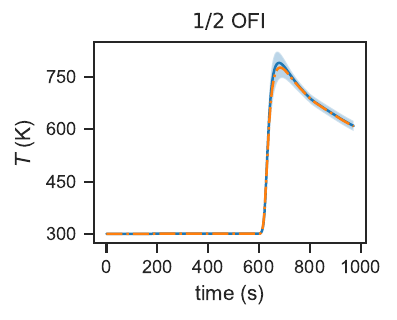}
    \end{subfigure}
    \begin{subfigure}[b]{0.3\textwidth}
        \centering
        \includegraphics[width=\textwidth]{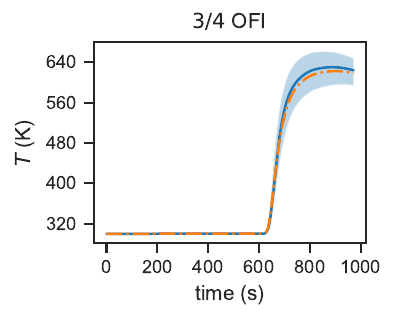}
    \end{subfigure}
    \begin{subfigure}[b]{0.3\textwidth}
        \centering
        \includegraphics[width=\textwidth]{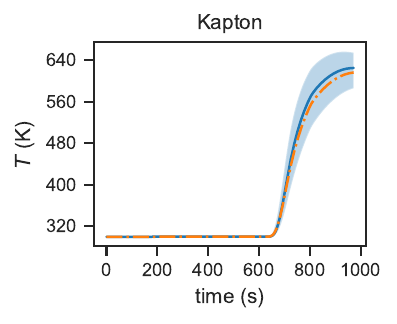}
    \end{subfigure}
    \caption{Temperature profile through the TPS at various points in the Mars entry scenario. At each location, we report the temperature prediction using the reference OFI properties along with the median and credible interval from the IFT-obtained properties.}
    \label{fig:marsuq}
\end{figure}

The Mars entry heating profile we use is based on data presented in~\cite{walker2015multifunctional}, which had used a preliminary MSL trajectory reconstruction data~\cite{bose2013initial} to generate cold wall heating profiles using the adjusted Sutton-Graves approximation~\cite{sutton1971general}.
The calculated heat flux and stagnation pressure are shown in Fig.~\ref{fig:marsdata}.
The adjusted values include the effect of vehicle radius, angle of attack, boundary layer radiation, and turbulence as determined using computational fluid dynamics calculations.
The Mars entry trajectory had been generated for a 4.5 m diameter MSL vehicle with a mass of 3257 kg, and an entry velocity of 5.9 km/s resulting in a peak cold wall heat flux of 70 W/cm2~\cite{walker2015multifunctional}.
This heating profile is used to size the OFI felt in a HIAD-like flexible TPS configuration, using ground truth and estimated thermal conductivities for OFI.
In total, the TPS was sized to be $25.4$ mm thick, which was chosen so that the temperature at the Kapton layer stays below the critical threshold of $625$ K.

\begin{table}[h!]
\centering
\begin{tabular}{| c c c c c c c |}
     \hline
     &  SiC & OFI-SiC & $1/4$ OFI & $1/2$ OFI & $3/4$ OFI & Kapton \\
     \hline\hline
     MSE & 0.612 & 4.07 & 48.3 & 40.9 & 57.3 & 109 \\
     Max err & 1.50 & 8.22 & 32.9 & 21.2 & 12.2 & 16.8\\
     \hline
\end{tabular}
\caption{Error between the IFT median prediction and reference temperature profile during the Mars entry of the TPS at evaluation points. The units are reported in Kelvin.}
\label{table:mars}
\end{table}

The results for this case are presented in Fig.~\ref{fig:marsuq}.
At each reported location in the TPS, we have good agreement between the ground truth and the IFT posterior median.
Observe that at the outer surface of the SiC fabric, there is no uncertainty: all thermal conductivities lead to the same profile.
Of course, this is to be expected, as there is essentially an imposed boundary condition.
In Table~\ref{table:mars}, we summarize the error between the IFT median prediction and the reference temperature profile of the TPS at the chosen evaluation points.
We report both the mean square error (MSE) and the maximum absolute error (Max err) across all timesteps.
Also we remark that as we predict further along the body, the uncertainty grows.
This is due to the uncertainty in the OFI thermal conductivity.
As the location moves further away from the boundary, this uncertainty naturally propagates through the temperature prediction.

\subsubsection{Earth entry simulations}

\begin{figure}
    \centering
    \begin{subfigure}[b]{0.33\textwidth}
        \centering
        \includegraphics[width = \textwidth]{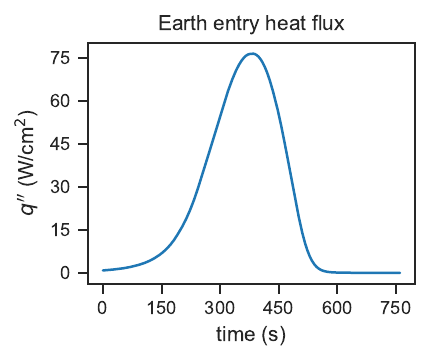}
        \caption{Earth heating profile}
    \end{subfigure}
    \begin{subfigure}[b]{0.33\textwidth}
        \centering
        \includegraphics[width = \textwidth]{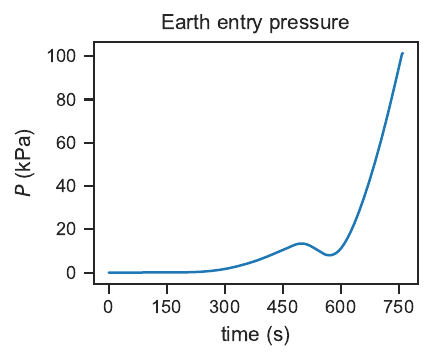}
        \caption{Earth stagnation pressure data}
    \end{subfigure}
    \caption{Cold wall heat flux and stagnation pressure used for sizing TPS for ISS return mission.}
    \label{fig:earthdata}
\end{figure}

The Earth entry heating profile we use is based on data presented in~\cite{walker2014preliminary}, which had generated a guided return trajectory for typical ISS return using the POST software~\cite{brauer1975program}.
The heat flux and stagnation pressure throughout this scenario are shown in Fig.~\ref{fig:earthdata}.
The ISS return trajectory had been generated for a 5 m diameter vehicle with a mass of 8223 kg, entry velocity of 7.5 km/s, and a lift to drag ratio of 0.1~\cite{walker2014preliminary}.
The trajectory velocity and density profiles had been used to calculate the cold wall heat flux profile using the Sutton-Graves approximation, resulting in a peak laminar convective cold wall heat flux of 75 W/cm$^2$~\cite{walker2014preliminary}.
Just as before, we use this heating profile to size the OFI felt in a HIAD-like flexible TPS configuration, using both the reference and IFT-derived thermal conductivities for OFI.
For Earth entry, we size the TPS to be $50.8$ mm thick.

\begin{figure}[htbp]
    \centering
    \begin{subfigure}[b]{0.3\textwidth}
        \centering
        \includegraphics[width=\textwidth]{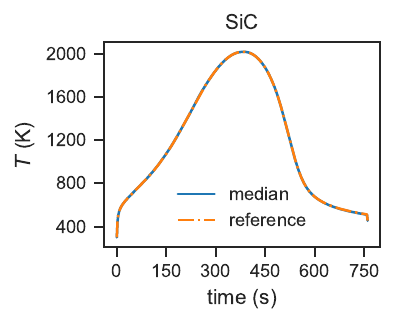}
    \end{subfigure}
    \begin{subfigure}[b]{0.3\textwidth}
        \centering
        \includegraphics[width=\textwidth]{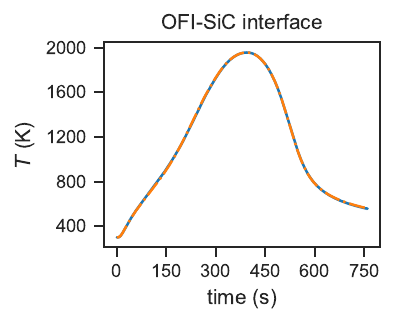}
    \end{subfigure}
    \begin{subfigure}[b]{0.3\textwidth}
        \centering
        \includegraphics[width=\textwidth]{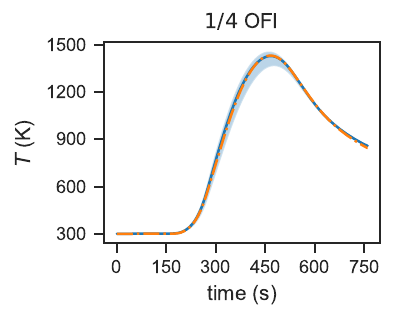}
    \end{subfigure}
    \hfill
    \begin{subfigure}[b]{0.3\textwidth}
        \centering
        \includegraphics[width=\textwidth]{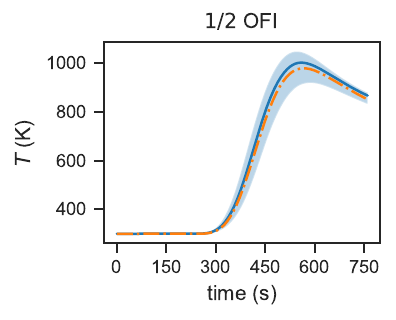}
    \end{subfigure}
    \begin{subfigure}[b]{0.3\textwidth}
        \centering
        \includegraphics[width=\textwidth]{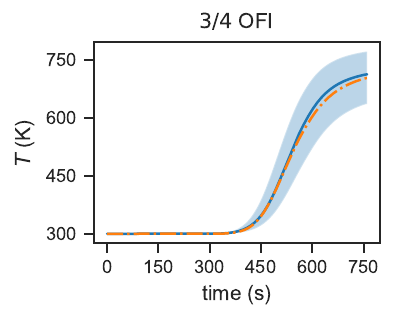}
    \end{subfigure}
    \begin{subfigure}[b]{0.3\textwidth}
        \centering
        \includegraphics[width=\textwidth]{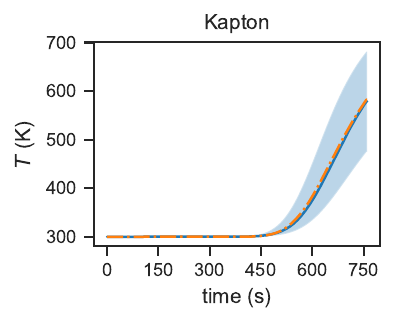}
    \end{subfigure}
    \caption{Temperature profile through the TPS at various points in the Earth entry scenario. At each location, we report the temperature prediction using the reference properties for OFI along with the median and credible interval from the IFT-derived properties.}
    \label{fig:earthuq}
\end{figure}

The resulting temperature profiles across the TPS for the Earth entry simulation are shown in Fig.~\ref{fig:earthuq}.
For the most part, the behavior is largely the same as it was for the Mars entry simulation, with the exception being a larger amount of uncertainty in the temperature prediction.
This is best evidenced at the bottom of the TPS (the Kapton layer) during the final timestep, where the uncertainty is highest.
The difference between the median and either end of the credible interval is roughly $30$ K for the Mars landing.
While for Earth, the same quantity is calculated to be just over $100$ K.
The reason for this can be attributed to the difference in atmospheric pressures in the two cases.
For higher pressures, the posterior over the thermal conductivity obtained from IFT was wider than in the low pressure experiments.
This results in a wider distribution after uncertainty propagation for Earth, as the pressure is much higher.
However, we still observe good agreement between the temperature with the reference properties and with the IFT median even with the larger discrepancies in the thermal conductivity reconstruction at high pressures.

\begin{table}[h!]
\centering
\begin{tabular}{| c c c c c c c |}
     \hline
     &  SiC & OFI-SiC & $1/4$ OFI & $1/2$ OFI & $3/4$ OFI & Kapton \\
     \hline\hline
     MSE & 0.949 & 12.7 & 80.9 & 313 & 66.3 & 15.5 \\
     Max err & 2.74 & 8.57 & 25.7 & 36.1 & 18.2 & 9.5\\
     \hline
\end{tabular}
\caption{Error between the IFT median prediction and reference temperature profile during the Earth entry scenario of the TPS at evaluation points. The units are reported in Kelvin.}
\label{table:earth}
\end{table}

Since the TPS experiences relatively lower pressures throughout most of the flight trajectory duration, the higher uncertainties in the IFT-estimated thermal conductivities at pressures closer to one atmosphere (as seen in Appendix~\ref{apdx:B}) do not significantly impact the in-depth median temperature predictions.
The uncertainty over the temperature increases with increasing distance from the outer surface of the TPS, but the median temperature predictions closely match the reference data predicted temperatures throughout the TPS.
The higher temperature uncertainties at the TPS cold side (Kapton film) would require TPS designers to increase the insulation thickness compared to the median IFT-predicted value to increase safety margins.

%% file: conclusions.tex
\section{Conclusions}

In this work, we demonstrated the ability of IFT for inferring the thermal conductivity of fibrous insulation from sparse experimental data, with direct application to spacecraft TPSs. 
By formulating conductivity inference as a Bayesian inverse problem constrained by the heat equation, we obtained posterior distributions that not only fuse available measurements with known physical laws, but also provide principled extrapolations into high-temperature regimes beyond the limits of ground testing.
As an example application, we made use of the resulting posterior distributions in an uncertainty propagation task for simulating spacecraft entry simulations.

When analyzing TPSs at temperatures above the available thermal properties limit, researchers typically either use linear extrapolation or keep the properties constant and equal to the available upper temperature data.
In either case, the resulting analysis can contain large errors, as thermal conductivity of high-porosity insulations is often a nonlinear function of temperature.
The outlined methodology using IFT has proven its capability to perform meaningful extrapolation of thermal properties that provide accurate median temperature predictions throughout the TPS thickness during the entry flight.
Mars entry was specifically chosen because of the lower pressure values experienced throughout the entry to show the best-case scenario.
On the other hand, the ISS return experiences higher pressures at the end of the trajectory and demonstrates the worst-case scenario.
The predicted temperature uncertainties at the TPS cold side were significant in both cases, especially for the ISS return case, but the median predicted temperatures closely matched temperature predictions using ground truth thermal conductivity property values.
This demonstrates the strength of the IFT approach when extrapolating properties beyond the upper limits of available data.

If the method were to be applied in its current state to sizing insulation for ground-based applications at atmospheric pressures, e.g. an industrial furnace, a higher fidelity heat transfer model would likely be necessary to ensure accurate predictions.
But for low-pressure applications such as designing insulation for on-orbit satellites, or for typical aerospace trajectories that see a wide range of pressures from vacuum to one atmosphere through the entry, the IFT predictions will result in accurate thermal modeling and TPS sizing.
Additional efforts as outlined in Appendix~\ref{apdx:B} and~\ref{apdx:valid} to improve the predictions at higher pressures and significantly decreasing uncertainties would be valuable. 
The most difficult part of modeling insulation thermal conductivity is obtaining the low-pressure ($0.001$ Torr) thermal conductivity which captures the contributions of radiation and solid conduction modes of heat transfer.
The current approach through IFT provides an excellent model of the low-pressure thermal conductivity over the extended temperature range.
Typically gas conduction contribution can be simply added to the low pressure thermal conductivity using standard methods if the effective insulation pore size (characteristic length for gas conduction) is known.
This method is used in~\cite{daryabeigi2024thermal} to obtain thermal conductivity of insulations at higher pressures from the low-pressure data, and is further addressed in Appendix~\ref{apdx:B}.

%% file: references.bib
@book{kaipio2006statistical,
  title={Statistical and computational inverse problems},
  author={Kaipio, Jari and Somersalo, Erkki},
  volume={160},
  year={2006},
  publisher={Springer Science \& Business Media}
}

@article{stuart2010inverse,
  title={Inverse problems: a Bayesian perspective},
  author={Stuart, Andrew M},
  journal={Acta numerica},
  volume={19},
  pages={451--559},
  year={2010},
  publisher={Cambridge University Press}
}

@article{alberts2023physics,
  title={Physics-informed information field theory for modeling physical systems with uncertainty quantification},
  author={Alberts, Alex and Bilionis, Ilias},
  journal={Journal of Computational Physics},
  volume={486},
  pages={112100},
  year={2023},
  publisher={Elsevier}
}

@article{ensslin2009information,
  title={Information field theory for cosmological perturbation reconstruction and nonlinear signal analysis},
  author={En{\ss}lin, Torsten A and Frommert, Mona and Kitaura, Francisco S},
  journal={Physical Review D—Particles, Fields, Gravitation, and Cosmology},
  volume={80},
  number={10},
  pages={105005},
  year={2009},
  publisher={APS}
}

@article{karniadakis2021physics,
  title={Physics-informed machine learning},
  author={Karniadakis, George Em and Kevrekidis, Ioannis G and Lu, Lu and Perdikaris, Paris and Wang, Sifan and Yang, Liu},
  journal={Nature Reviews Physics},
  volume={3},
  number={6},
  pages={422--440},
  year={2021},
  publisher={Nature Publishing Group}
}

@book{cartier2006functional,
  title={Functional integration: action and symmetries},
  author={Cartier, Pierre and DeWitt-Morette, C{\'e}cile},
  year={2006},
  publisher={Cambridge University Press}
}

@book{feynman2010quantum,
  title={Quantum mechanics and path integrals},
  author={Feynman, Richard P and Hibbs, Albert R and Styer, Daniel F},
  year={2010},
  publisher={Courier Corporation}
}

@book{albeverio1976mathematical,
  title={Mathematical theory of Feynman path integrals},
  author={Albeverio, Sergio and H{\"o}egh-Krohn, Raphael and Mazzucchi, Sonia},
  volume={523},
  year={1976},
  publisher={Springer}
}

@article{parisi1988statistical,
  title={Statistical field theory},
  author={Parisi, Giorgio and Shankar, Ramamurti},
  year={1988},
  publisher={Westview Press}
}

@book{glimm2012quantum,
  title={Quantum physics: a functional integral point of view},
  author={Glimm, James and Jaffe, Arthur},
  year={2012},
  publisher={Springer Science \& Business Media}
}

@article{bui2013computational,
  title={A computational framework for infinite-dimensional Bayesian inverse problems Part I: The linearized case, with application to global seismic inversion},
  author={Bui-Thanh, Tan and Ghattas, Omar and Martin, James and Stadler, Georg},
  journal={SIAM Journal on Scientific Computing},
  volume={35},
  number={6},
  pages={A2494--A2523},
  year={2013},
  publisher={SIAM}
}

@article{chen2021solving,
  title={Solving and learning nonlinear PDEs with Gaussian processes},
  author={Chen, Yifan and Hosseini, Bamdad and Owhadi, Houman and Stuart, Andrew M},
  journal={Journal of Computational Physics},
  volume={447},
  pages={110668},
  year={2021},
  publisher={Elsevier}
}

@article{bai2024gaussian,
  title={Gaussian processes for Bayesian inverse problems associated with linear partial differential equations},
  author={Bai, Tianming and Teckentrup, Aretha L and Zygalakis, Konstantinos C},
  journal={Statistics and Computing},
  volume={34},
  number={4},
  pages={139},
  year={2024},
  publisher={Springer}
}

@article{cotter2010approximation,
  title={Approximation of Bayesian inverse problems for PDEs},
  author={Cotter, Simon L and Dashti, Masoumeh and Stuart, Andrew M},
  journal={SIAM journal on numerical analysis},
  volume={48},
  number={1},
  pages={322--345},
  year={2010},
  publisher={SIAM}
}

@article{yang2021b,
  title={B-PINNs: Bayesian physics-informed neural networks for forward and inverse PDE problems with noisy data},
  author={Yang, Liu and Meng, Xuhui and Karniadakis, George Em},
  journal={Journal of Computational Physics},
  volume={425},
  pages={109913},
  year={2021},
  publisher={Elsevier}
}

@article{alberts2024well,
  title={On the well-posedness of inverse problems under information field theory: Application to model-form error detection},
  author={Alberts, Alex and Bilionis, Ilias},
  journal={arXiv preprint arXiv:2401.14224},
  year={2024}
}

@inproceedings{welling2011bayesian,
  title={Bayesian learning via stochastic gradient Langevin dynamics},
  author={Welling, Max and Teh, Yee W},
  booktitle={Proceedings of the 28th international conference on machine learning (ICML-11)},
  pages={681--688},
  year={2011},
  organization={Citeseer}
}

@article{cotter2013mcmc,
  title={MCMC methods for functions: modifying old algorithms to make them faster},
  author={Cotter, Simon L and Roberts, Gareth O and Stuart, Andrew M and White, David},
  year={2013}
}

@article{ghattas2021learning,
  title={Learning physics-based models from data: perspectives from inverse problems and model reduction},
  author={Ghattas, Omar and Willcox, Karen},
  journal={Acta Numerica},
  volume={30},
  pages={445--554},
  year={2021},
  publisher={Cambridge University Press}
}

@article{williams1995gaussian,
  title={Gaussian processes for regression},
  author={Williams, Christopher and Rasmussen, Carl},
  journal={Advances in neural information processing systems},
  volume={8},
  year={1995}
}

@article{hao2025neural,
  title={Neural information field filter},
  author={Hao, Kairui and Bilionis, Ilias},
  journal={Mechanical Systems and Signal Processing},
  volume={226},
  pages={112253},
  year={2025},
  publisher={Elsevier}
}

@book{williams2006gaussian,
  title={Gaussian processes for machine learning},
  author={Williams, Christopher KI and Rasmussen, Carl Edward},
  volume={2},
  number={3},
  year={2006},
  publisher={MIT press Cambridge, MA}
}

@article{zhu1997gaussian,
  title={Gaussian regression and optimal finite dimensional linear models},
  author={Zhu, Huaiyu and Williams, Christopher KI and Rohwer, Richard and Morciniec, Michal},
  year={1997},
  publisher={Aston University}
}

@article{quinonero2005unifying,
  title={A unifying view of sparse approximate Gaussian process regression},
  author={Quinonero-Candela, Joaquin and Rasmussen, Carl Edward},
  journal={Journal of machine learning research},
  volume={6},
  number={Dec},
  pages={1939--1959},
  year={2005}
}

@article{kingma2014adam,
  title={Adam: A method for stochastic optimization},
  author={Kingma, Diederik P},
  journal={arXiv preprint arXiv:1412.6980},
  year={2014}
}

@software{deepmind2020jax,
  title = {The {D}eep{M}ind {JAX} {E}cosystem},
  author = {DeepMind and Babuschkin, Igor and Baumli, Kate and Bell, Alison and Bhupatiraju, Surya and Bruce, Jake and Buchlovsky, Peter and Budden, David and Cai, Trevor and Clark, Aidan and Danihelka, Ivo and Dedieu, Antoine and Fantacci, Claudio and Godwin, Jonathan and Jones, Chris and Hemsley, Ross and Hennigan, Tom and Hessel, Matteo and Hou, Shaobo and Kapturowski, Steven and Keck, Thomas and Kemaev, Iurii and King, Michael and Kunesch, Markus and Martens, Lena and Merzic, Hamza and Mikulik, Vladimir and Norman, Tamara and Papamakarios, George and Quan, John and Ring, Roman and Ruiz, Francisco and Sanchez, Alvaro and Sartran, Laurent and Schneider, Rosalia and Sezener, Eren and Spencer, Stephen and Srinivasan, Srivatsan and Stanojevi\'{c}, Milo\v{s} and Stokowiec, Wojciech and Wang, Luyu and Zhou, Guangyao and Viola, Fabio},
  url = {http://github.com/google-deepmind},
  year = {2020},
}

@article{robbins1951stochastic,
  title={A stochastic approximation method},
  author={Robbins, Herbert and Monro, Sutton},
  journal={The annals of mathematical statistics},
  pages={400--407},
  year={1951},
  publisher={JSTOR}
}

@techreport{pavlosky1974apollo,
  title={Apollo experience report: Thermal protection subsystem},
  author={Pavlosky, James E and others},
  year={1974}
}

@book{curry1993space,
  title={Space shuttle orbiter thermal protection system design and flight experience},
  author={Curry, Donald M},
  volume={104773},
  year={1993},
  publisher={Lyndon B. Johnson Space Center}
}

@inproceedings{olds2013irve,
  title={IRVE-3 post-flight reconstruction},
  author={Olds, Aaron and Beck, Roger and Bose, David M and White, Joseph and Edquist, Karl T and Hollis, Brian R and Lindell, Michael and Cheatwood, FM and Gsell, Valerie and Bowden, Ernest L},
  booktitle={AIAA Aerodynamic Decelerator Systems (ADS) Conference},
  pages={1390},
  year={2013}
}

@inproceedings{dinonno2024low,
  title={Low-earth orbit flight test of an inflatable decelerator (loftid) mission overview and science return},
  author={DiNonno, John and Cheatwood, Neil},
  booktitle={AIAA SCITECH 2024 Forum},
  pages={1309},
  volume = {1390},
  year={2024}
}

@article{flynn1969radial,
  title={A radial-flow apparatus for determining the thermal conductivity of loose-fill insulations to high temperatures},
  author={Flynn, DR},
  journal={Precision Measurement and Calibration: Heat},
  volume={300},
  pages={325},
  year={1969},
  publisher={Department of Commerce, National Bureau of Standards}
}

@article{lee2019radiation,
  title={Radiation heat transfer through carbon fiber materials: Experiment vs theory},
  author={Lee, Siu-Chun},
  journal={Journal of Thermophysics and Heat Transfer},
  volume={33},
  number={2},
  pages={370--377},
  year={2019},
  publisher={American Institute of Aeronautics and Astronautics}
}

@article{zarr2020nist,
  title={NIST-NPL bilateral comparison of guarded-hot-plate laboratories from 20° C to 160° C},
  author={Zarr, Robert R and Wu, Jiyu and Liu, Hung-Kung},
  journal={NIST, Gaithersburg, MD, USA, Tech. Note},
  volume={2059},
  year={2020}
}

@techreport{daryabeigi2024thermal,
  title={Thermal Modeling and Testing of High-Temperature Refractory Ceramic Insulation Felts},
  author={Daryabeigi, Kamran},
  year={2024},
  number = {20240003609},
  institution = {NASA Langley Research Center}
}

@standard{ASTM:C177-19,
  title        = {Standard Test Method for Steady‑State Heat Flux Measurements and Thermal Transmission Properties by Means of the Guarded‑Hot‑Plate Apparatus},
  author  = {ASTM International},
  address      = {West Conshohocken, PA},
  number       = {C177‑19},
  year         = {2019},
  month        = {January},
  note         = {},
}

@standard{ASTM:C518,
  title        = {Standard Test Method for Steady-State Thermal Transmission Properties by Means of the Heat Flow Meter Apparatus},
  author  =    {ASTM International},
  address      = {West Conshohocken, PA},
  number       = {ASTM Standard C518},
  year         = {2023},
  month        = {Feburary},
  note         = {},
}

@article{gembarovic2007method,
  title={A method for thermal diffusivity determination of thermal insulators},
  author={Gembarovic, Jozef and Taylor, Raymond E},
  journal={International Journal of Thermophysics},
  volume={28},
  number={6},
  pages={2164--2175},
  year={2007},
  publisher={Springer}
}

@standard{ASTM:E1269,
  title        = {Standard Test Method for Determining Specific Heat Capacity by Differential Scanning Calorimetry},
  author  = {ASTM International},
  address      = {West Conshohocken, PA},
  number       = {E1269-11},
  year         = {2011}
}

@article{daryabeigi2011combined,
  title={Combined heat transfer in high-porosity high-temperature fibrous insulation: Theory and experimental validation},
  author={Daryabeigi, Kamran and Cunnington, George R and Knutson, Jeffrey R},
  journal={Journal of thermophysics and heat transfer},
  volume={25},
  number={4},
  pages={536--546},
  year={2011}
}

@inproceedings{edquist2009aerothermodynamic,
  title={Aerothermodynamic design of the Mars science laboratory heatshield},
  author={Edquist, Karl and Dyakonov, Artem and Wright, Michael and Tang, Chun},
  booktitle={41st AIAA Thermophysics Conference},
  pages={4075},
  year={2009}
}

@article{wright2014sizing,
  title={Sizing and margins assessment of mars science laboratory aeroshell thermal protection system},
  author={Wright, Michael J and Beck, Robin AS and Edquist, Karl T and Driver, David and Sepka, Steven A and Slimko, Eric M and Willcockson, William H},
  journal={Journal of Spacecraft and Rockets},
  volume={51},
  number={4},
  pages={1125--1138},
  year={2014},
  publisher={American Institute of Aeronautics and Astronautics}
}

@inproceedings{walker2015multifunctional,
  title={A multifunctional hot structure heat shield concept for planetary entry},
  author={Walker, Sandra P and Daryabeigi, Kamran and Samareh, Jamshid A and Wagner, Robert and Waters, William},
  booktitle={20th AIAA International Space Planes and Hypersonic Systems and Technologies Conference},
  pages={3530},
  year={2015}
}

@inproceedings{walker2014preliminary,
  title={Preliminary development of a multifunctional hot structure heat shield},
  author={Walker, Sandra P and Daryabeigi, Kamran and Samareh, Jamshid A and Armand, Sasan C},
  booktitle={55th AIAA/ASMe/ASCE/AHS/SC Structures, Structural Dynamics, and Materials Conference},
  pages={0350},
  year={2014}
}

@inproceedings{bose2013initial,
  title={Initial assessment of Mars Science Laboratory heatshield instrumentation and flight data},
  author={Bose, Deepak and Olson, Michael and Laub, Bernard and White, Todd and Feldman, Jay and Santos, Jose and Mahzari, Milad and MacLean, Matthew and Dufrene, Aaron and Holden, Michael},
  booktitle={51st AIAA Aerospace Sciences Meeting including the New Horizons Forum and Aerospace Exposition},
  pages={908},
  volume = {0908},
  year={2013}
}

@techreport{sutton1971general,
  title={A general stagnation-point convective heating equation for arbitrary gas mixtures},
  author={Sutton, Kenneth and Graves Jr, Randolph A},
  number = {NASA TN D-7564},
  year={1971}
}

@techreport{brauer1975program,
  title={Program to optimize simulated trajectories (POST). Volume 1: Formulation manual},
  author={Brauer, GL and Cornick, DE and Habeger, AR and Petersen, FM and Stevenson, R},
  year={1975}
}

@misc{kratos_thermal_testing,
  author       = {{Kratos Defense \& Security Solutions, Inc.}},
  title        = {Thermal Testing of Materials},
  howpublished = {\url{https://www.kratosdefense.com/about/divisions/defense-and-rocket-support-services/sre/thermal-testing-of-materials}},
  year         = {n.d.},
}

@book{kennard1938kinetic,
  title={Kinetic theory of gases},
  author={Kennard, Earle H and others},
  volume={483},
  year={1938},
  publisher={McGraw-hill New York}
}

@incollection{marcussen1985thermal,
  author    = {Marcussen, L.},
  title     = {Prediction of Effective Thermal Conductivity for Fibrous Media},
  booktitle = {Thermal Conductivity},
  editor    = {Yarbrough, David},
  volume    = {19},
  year      = {1985},
  publisher = {Plenum Press},
  pages     = {75--84}
}

@article{yu2018deep,
  title={The deep Ritz method: a deep learning-based numerical algorithm for solving variational problems},
  author={Yu, Bing and others},
  journal={Communications in Mathematics and Statistics},
  volume={6},
  number={1},
  pages={1--12},
  year={2018},
  publisher={Springer}
}
